\documentstyle[pre,aps,epsfig,amssymb]{revtex}

\def\YHC#1#2#3#4#5#6{\setlength{\unitlength}{#6}
\makebox{
\begin{picture}(23,45)(-3,-25)
\thicklines
\put(8.335,1){\line(0,1){9}}
\put(8.335,10){\line(3,2){8}}
\put(8.335,10){\line(-3,2){8}}
\put(8.3335,-1){\makebox(0,0)[t]{$#3$}}
\put(8.3335,-11){\makebox(0,0)[t]{\scriptsize $#4$}}
\put(8.3335,-21){\makebox(0,0)[t]{\scriptsize $#5$}}
\put(17.167,16.667){\makebox(0,0)[l]{$#2$}}
\put(-.5,16.667){\makebox(0,0)[r]{$#1$}}
\end{picture}}}

\def\laHC#1#2#3#4#5#6{\setlength{\unitlength}{#6}
\makebox{
\begin{picture}(23,45)(-3,-25)
\thicklines
\put(8.335,12.667){\line(0,-1){9}}
\put(8.335,3.667){\line(3,-2){8}}
\put(8.335,3.667){\line(-3,-2){8}}
\put(-.5,-3){\makebox(0,0)[r]{$#1$}}
\put(17.167,-3){\makebox(0,0)[l]{$#3$}}
\put(8.335,14.667){\makebox(0,0)[b]{$#2$}}
\put(8.335,-11){\makebox(0,0)[t]{\scriptsize $#4$}}
\put(8.335,-21){\makebox(0,0)[t]{\scriptsize $#5$}}
\end{picture}}}

\def\Tup#1#2#3#4{\setlength{\unitlength}{#4}
\makebox{
\begin{picture}(11,13)(-.5,-3)
\thicklines
\put(0,0){\line(1,0){10}}
\put(0,0){\line(3,5){5}}
\put(10,0){\line(-3,5){5}}
\put(5,-.51){\makebox(0,0)[t]{$#3$}}
\put(8.5,5.5){\makebox(0,0)[l]{$#2$}}
\put(1.5,5.5){\makebox(0,0)[r]{$#1$}}
\end{picture}}}

\def\Tdown#1#2#3#4{\setlength{\unitlength}{#4}
\makebox{
\begin{picture}(11,13)(-.5,0)
\thicklines
\put(0,10){\line(1,0){10}}
\put(0,10){\line(3,-5){5}}
\put(10,10){\line(-3,-5){5}}
\put(5,11){\makebox(0,0)[b]{$#2$}}
\put(1.5,5){\makebox(0,0)[r]{$#1$}}
\put(8.5,5){\makebox(0,0)[l]{$#3$}}
\end{picture}}}

\begin{document}
\draft
\title{On the integrability of the square-triangle random tiling
  model}
\author{Jan de Gier \and Bernard Nienhuis}
\address{Instituut voor Theoretische Fysica, Universiteit van
  Amsterdam, Valckenierstraat 65, 1018 XE Amsterdam, The Netherlands}
\date{\today}
\maketitle
\begin{abstract}
It is shown that the square-triangle random tiling model is equivalent
to an asymmetric limit of the three coloring model on the honeycomb
lattice. The latter model is known to be the $O(n)$ model at $T=0$ and
corresponds to the integrable model connected to the affine
$A_2^{(1)}$ Lie algebra. Thus it is shown that the weights of the
square-triangle random tiling satisfy the Yang-Baxter equation, albeit
in a singular limit of a more general model. The three coloring model
for general vertex weights is solved by algebraic Bethe Ansatz.
\end{abstract}

\section{Introduction}
Random tiling models have gained renewed interest in the past years by
the discovery of quasi-crystals \cite{Shechtman:1984}. They provide an
example of the entropic
stability of structures whose diffraction pattern has a rotational
symmetry which is incompatible with periodicity. As such they offer an
explanation of the existence of quasi-crystalline alloys \cite{Henley:1991}.
In analogy with the diamond covering used to describe the ground state
configurations of the triangular Ising anti-ferromagnet
\cite{Blote:1982}, random tilings can be described by a domain wall
structure. The difficulty is that there is more than one type of
domain wall, in contrast to the diamond covering. This fact makes it
much more difficult to solve the model by coordinate Bethe Ansatz, as was tried
for the octagonal square-rhombus tiling \cite{Li:1992}. Widom
\cite{Widom:1993} however succeeded in diagonalizing the transfer
matrix using Bethe Ansatz for the 12-fold square-triangle random tiling
model. Shortly after that Kalugin \cite{Kalugin:1994} was able to find
a closed expression for the entropy as a function of the domain wall
densities in part of the phase diagram. More recently a solution of an
octagonal random tiling has also been found \cite{Gier:1996a}. 

Being solvable by coordinate Bethe Ansatz, it is natural to look for
solutions of the Yang-Baxter equation \cite{Baxter:1982} for these
models. This would provide an answer to why these models are
integrable and a canonical way to diagonalize the transfer matrix via
the algebraic Bethe Ansatz \cite{Takhtadzhan:1984}. The matrix of
Boltzmann weights \cite{Baxter:1982} however, is not  invertible,
which is necessary to obtain commutativity of the transfer matrix. In
this paper we show that the square-triangle random tiling model is a
singular limit case of a more general model obeying the Yang-Baxter equation. 

\section{The model}
We shall consider a vertex model on the square lattice, whose
Boltzmann weights are denoted by
\begin{equation}
W(\mu,\alpha;\beta,\nu|u) = 
\setlength{\unitlength}{.8mm}
\vcenter{\hbox{
\begin{picture}(28,28)(-4,-14)
\linethickness{1pt}
\put(10,-10){\line(0,1){20}}
\put(0,0){\line(1,0){20}}
\put(-1,0){\makebox(0,0)[r]{$\mu$}}
\put(10,-11){\makebox(0,0)[t]{$\alpha$}}
\put(10,11){\makebox(0,0)[b]{$\beta$}}
\put(21,0){\makebox(0,0)[l]{$\nu$}}
\end{picture}}}
\end{equation}
Each edge of the lattice can be in one of three different states,
1,2 and 3. The partition function of the model is
given by 
\begin{equation}
Z = \sum_{\rm config.} \prod_i W(i),\label{Zdef}
\end{equation}
where we sum over all configurations which are weighted by the product
over the vertices $i$ of their local Boltzmann weights $W(i)$. The
explicit form of the 
weights can be found in Table~\ref{ta:BoltzW}.
The model can be written in terms of the weights
$W_0(\mu,\alpha;\beta,\nu)$ associated with the affine Lie algebra
$A_2^{(1)}$, which can be found in \cite{Chudnovsky:1980,Cherednik:1980,Perk:1981}. 
\begin{equation}
W(\mu,\alpha;\beta,\nu|u)=O_{\mu\mu'}W_0(\mu',\alpha;\beta,\nu'|u)O_{\nu'\nu},
\end{equation}
where $O={\rm diag} \{x_1,x_2,x_3\}$.
The weights $W_0(\alpha,\nu;\beta,\nu)$ satisfy the star-triangle or
Yang-Baxter equation (YBE) \cite{Baxter:1982}, 
\begin{eqnarray}
\displaystyle \sum_{\gamma,\mu'',\nu''} W_0(\mu,\nu;\nu'',\mu''|v-u)
W_0(\mu'',\alpha;\gamma,\mu'|v) W_0(\nu'',\gamma;\beta,\nu'|u)=\nonumber\\
\displaystyle \sum_{\gamma,\mu'',\nu''} W_0(\nu,\alpha;\gamma,\nu''|u)
W_0(\mu,\gamma;\beta,\mu''|v) W_0(\mu'',\nu'';\nu',\mu'|v-u).\label{YBE1}
\end{eqnarray}
Equation~(\ref{YBE1}) can be written more
elegantly by defining the operators $L(u)$ 
\begin{equation}
\begin{array}{l}
\displaystyle \left(L(u)_{\mu\nu}\right)_{\alpha\beta} = W_0(\mu,\alpha;\beta,\nu|u),\\[2mm]
\displaystyle L_a(u):\;V_a\otimes {\Bbb C}^3 \rightarrow V_a\otimes {\Bbb C}^3,\;V_a\simeq {\Bbb C}^3.\label{Laxdef}
\end{array}
\end{equation}
The auxiliary label $a$ is introduced for later convenience. We shall
omit it if there is no confusion on which space $L(u)$ acts. In this
language the vertex states $\alpha$ $(\alpha=1,2,3)$ are represented
by the standard basis $e_\alpha$ of ${\Bbb C}^3$. $V_a\simeq
V_b\simeq{\Bbb C}^3$ are so-called auxiliary 
spaces, corresponding to the horizontal edges of a vertex. 
The transfer matrix ${\cal T}=\sum_{\mu=1}^3 T(u)_{\mu\mu}$ on a
lattice of horizontal size $N$ can be written in
terms of the local operators $L_a(u)$~(\ref{Laxdef}) as
\begin{equation}
\begin{array}{l} 
\displaystyle T_a(u) = \left(\prod_{j=1}^N O_a^2 L_a^j(u)\right),\\[2mm]
\displaystyle T_a(u): V_a\otimes{\cal H}\rightarrow V_a\otimes{\cal H},\;{\cal
  H}=\bigotimes_{j=1}^N {\Bbb C}^3,\\[2mm]
\displaystyle {\cal T}(u)={\rm Tr}_a T_a(u).
\end{array}
\label{eq:monodef}
\end{equation}
The trace is taken only over the auxiliary space matrix structure.
The operators $L_a^j(u)$ act as $L_a(u)$ on the $j$th factor
in ${\cal H}$ and as the identity on all other factors,
\begin{equation}
L^j(u)_{\mu\nu} = \stackrel{\rm j-1\;
  times}{\overbrace{I\otimes\cdots\otimes I\otimes}}
L(u)_{\mu\nu}\otimes\stackrel{\rm N-j\; times}{\overbrace{
I\otimes\cdots\otimes I}},
\end{equation}
where $I$ is the identity on ${\Bbb C}^3$. The partition function
(\ref{Zdef}) on a lattice of size $N\times M$ can then be written as
\begin{equation}
Z = {\rm Tr}_{\cal H}{\cal T}(u)^M.\label{partsum}
\end{equation}
Furthermore we define the $R$-matrix as
\begin{equation}
\begin{array}{l}
\displaystyle \left(R(v-u)_{\mu\nu}\right)_{\alpha\beta} = W_0(\mu,\alpha;\beta,\nu|v-u),\\[2mm]
\displaystyle R_{ab}(v-u):\;V_a\otimes V_b\rightarrow V_a\otimes V_b.
\end{array}\label{eq:Rdef}
\end{equation}
Here too, the Roman labels $a$ and $b$ indicate on which auxiliary space $R$
is acting. Greek labels will be used to indicate matrix elements.
The YBE (\ref{YBE1}) can be written as
\begin{eqnarray}
(R(v-u)_{\mu\mu''})_{\nu\nu''}\left[L(v)_{\mu''\mu'}L(u)_{\nu''\nu'}\right]
= \nonumber\\
\left[L(u)_{\nu\nu''}L(v)_{\mu\mu''}\right](R(v-u)_{\mu''\mu'})_{\nu''\nu'},
\label{YBE2}\end{eqnarray}
where summation over repeated indices is understood. Each element
$L(u)_{\mu\nu}$ of $L(u)$ is an operator acting on
${\Bbb C}^3$. We shall regard $R(u-v)$ as a $9\times 9$ matrix which
is given explicitly in Table~\ref{ta:BoltzW}.
In a compact notation the YBE (\ref{YBE2}) can be written as an
operator equation on the tensor product $V_a\otimes V_b\otimes
{\Bbb C}^3$. It then assumes the guise 
\begin{equation}
R_{ab}(v-u)\cdot\left[L_a(v)\otimes L_b(u)\right] =
\left[L_b(u)\otimes L_a(v)\right]\cdot R_{ab}(v-u).\label{YBE3}
\end{equation}
From (\ref{YBE3}) we obtain the YBE for the matrix $T(u)$ as defined
in (\ref{eq:monodef}). 
\begin{equation}
R_{ab}(v-u)\cdot\left[T_a(v)\otimes T_b(u)\right] =
\left[T_b(u)\otimes T_a(v)\right]\cdot R_{ab}(v-u).\label{YBET}
\end{equation}
From the expression (\ref{partsum}) for the partition sum one sees that
the leading term is given by the largest eigenvalue of ${\cal T}$. For
the sake of completeness we give the explicit diagonalization of the
transfer matrix using the algebraic Bethe Ansatz in appendix
\ref{diagT}. This is just a special case of the trigonometric case of
Kulish and Reshetikhin \cite{Kulish:1983}.
\begin{table}[h]
\begin{eqnarray}
&&W(\mu,\mu;\mu,\mu|u)=x_\mu^2\sinh (u+\lambda),\nonumber\\[2mm]
&&W(\mu,\nu;\mu,\nu|u)=x_\mu x_\nu {\rm e}^{-u\;{\rm sgn} (\mu-\nu)}
\sinh(\lambda), \nonumber\\[2mm]
&&W(1,2;2,1|u)=x_1^2y_3^{-2}\sinh (u),\;
W(1,3;3,1|u)=x_1^2y_2^{-2}\sinh (u),\nonumber\\[2mm] 
&&W(2,3;3,2|u)=x_2^2y_1^{-2}\sinh (u),\;
W(2,1;1,2|u)=x_2^2y_3^2\sinh (u),\nonumber\\[2mm] 
&&W(3,1;1,3|u)=x_3^2y_2^2\sinh (u),\;
W(3,2;2,3|u)=x_3^2y_1^2\sinh (u),\nonumber\\[2mm]
&&\begin{array}{l}
R(u)=
\left(
\begin{array}{@{}ccccccccc}
s_2 & 0 & 0 & 0 & 0 & 0 & 0 & 0 & 0\\
0 & y_3^{-2}s_1 & 0 & {\rm e}^{u}s_0 & 0 & 0 & 0 & 0 & 0\\
0 & 0 & y_2^{-2}s_1 & 0 & 0 & 0 & {\rm e}^{u}s_0 & 0 & 0\\
0 & {\rm e}^{-u}s_0 & 0 & y_3^2s_1 & 0 & 0 & 0 & 0 & 0\\
0 & 0 & 0 & 0 & s_2 & 0 & 0 & 0 & 0\\
0 & 0 & 0 & 0 & 0 & y_1^{-2}s_1 & 0 & {\rm e}^{u}s_0 & 0 \\
0 & 0 & {\rm e}^{-u}s_0 & 0 & 0 & 0 & y_2^2s_1 & 0 & 0\\
0 & 0 & 0 & 0 & 0 & {\rm e}^{-u}s_0 & 0 & y_1^2s_1 & 0\\
0 & 0 & 0 & 0 & 0 & 0 & 0 & 0 & s_2
\end{array}\;\right).\end{array}\nonumber
\end{eqnarray}
\caption{Boltzmann weights $W$ and $R$-matrix corresponding to the
  weights $W_0$. Here we use the abbreviations $s_0=\sinh(\lambda)$,
  $s_1=\sinh(u)$, $s_2=\sinh(u+\lambda)$.}
\label{ta:BoltzW} 
\end{table}   
\section{Honeycomb lattice}
As was already shown by Reshetikhin \cite{Reshetikhin:1989}, at a
special value of the spectral parameter $u$, the model defined by the
weights $W_0$ factorizes on the honeycomb lattice. In this section we
rederive this result for the model in Table~\ref{ta:BoltzW}. Consider
the operators 
\begin{equation}
OL(u)O\tau:{\Bbb C}^3\otimes V\rightarrow V\otimes{\Bbb C}^3,
\end{equation}
where $\tau$ is the permutation operator in ${\Bbb C}^3\otimes
{\Bbb C}^3$ (recall that $V\simeq{\Bbb C}^3)$, $\tau (e_\alpha\otimes
e_\beta)=e_\beta\otimes e_\alpha$. The eigenvectors of $OL(u)O\tau$
are given by 
\begin{equation}
\begin{array}{@{}l}
\begin{array}{@{}l}
\displaystyle \bar{e}_\gamma={1\over\sqrt{1+{\rm e}^{-2\lambda}}}(x_\beta y_\gamma
e_\beta\otimes e_\alpha-y_\gamma^{-1}x_\alpha{\rm e}^{-\lambda}e_\alpha\otimes
e_\beta),\\[4mm] 
\displaystyle \bar{f}_{(\beta\alpha)}={1\over\sqrt{1+{\rm e}^{2\lambda}}}(y_\gamma x_\beta
e_\beta\otimes e_\alpha + y_\gamma^{-1}x_\alpha
{\rm e}^{\lambda}e_\alpha\otimes e_\beta), 
\end{array}
\\[4mm]
\displaystyle\bar{f}_{(\gamma\gamma)} =
\vphantom{{1\over\sqrt{1+{\rm e}^{2\lambda}}}}e_\gamma\otimes e_\gamma ,
\end{array}\end{equation}
with $(\alpha,\beta,\gamma)$ a cyclic permutation of (1,2,3).
These vectors satisfy the eigenvalue equations
\begin{equation}
\begin{array}{l}
\displaystyle OL(u)O\tau \bar{e}_\gamma = -x_\alpha x_\beta\sinh (u-\lambda)
\bar{e}_\gamma,\\[2mm]  
\displaystyle OL(u)O\tau \bar{f}_{(\mu\nu)} = x_\mu x_\nu\sinh
(u+\lambda)\bar{f}_{(\mu\nu)}.\\[2mm]
\end{array}\end{equation}
Thus at $u=-\lambda$ the operator $L\tau$ becomes a
projector. 
\begin{figure}[h]
\setlength{\unitlength}{.8mm}
\centerline{
\begin{picture}(100,40)(-2,-10)
\linethickness{1pt}
\put(0,0){\line(1,1){20}}
\put(0,-1){\makebox(0,0)[t]{$\mu$}}
\put(20,21){\makebox(0,0)[b]{$\nu$}}
\put(0,20){\line(1,-1){20}}
\put(0,21){\makebox(0,0)[b]{$\beta$}}
\put(20,-1){\makebox(0,0)[t]{$\alpha$}}
\put(30,10){\vector(1,0){10}}
\put(50,-7.0720678){\line(1,1){10}}
\put(50,-8.0720678){\makebox(0,0)[t]{$\mu$}}
\put(70,-7.0720678){\line(-1,1){10}}
\put(70,-8.0720678){\makebox(0,0)[t]{$\alpha$}}
\put(60,3.0720678){\line(0,1){14.142136}}
\put(62,10.144136){\makebox(0,0)[l]{$\sinh (2\lambda)\Sigma_\gamma$}}
\put(50,27.21420){\line(1,-1){10}}
\put(50,28.21420){\makebox(0,0)[b]{$\beta$}}
\put(70,27.21420){\line(-1,-1){10}}
\put(70,28.21420){\makebox(0,0)[b]{$\nu$}}
\end{picture}}
\vskip5mm
\caption{Factorization of the weights}
\label{fig:sq2hc}
\end{figure}
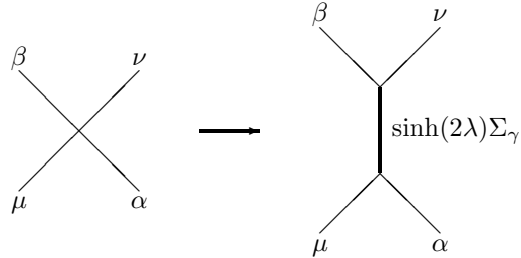
Introducing the dual vectors $\bar{e}^*_\gamma$ with the properties
\begin{equation}
\bar{e}^*_\gamma\cdot \bar{e}_{\gamma'}^{\vphantom{*}} = x_\alpha
x_\beta\delta_{\gamma\gamma'},
\end{equation}
we can write $OLO\tau$ at $u=-\lambda$ as
\begin{equation}
OL(-\lambda)O\tau = \sinh (2\lambda)\sum_{\gamma=1}^3
\bar{e}_\gamma\bar{e}_\gamma^*. 
\end{equation}
Give the y-vertices of the honeycomb lattice the weights
$\bar{e}^*_\gamma\cdot (e_\beta\otimes e_\nu)$ and the $\lambda$-vertices the
weights $(e_\mu\otimes e_\alpha)\cdot \bar{e}_\gamma$. Graphically,
the matrix element $(L(-\lambda)\tau_{\mu\beta})_{\alpha\nu}$,
corresponding to the vertex weight $W(\mu,\alpha;\beta,\nu|-\lambda)$,
can be written as a sum over products of two vertices of the honeycomb
lattice, see Figure~\ref{fig:sq2hc}.

\begin{figure}[h]
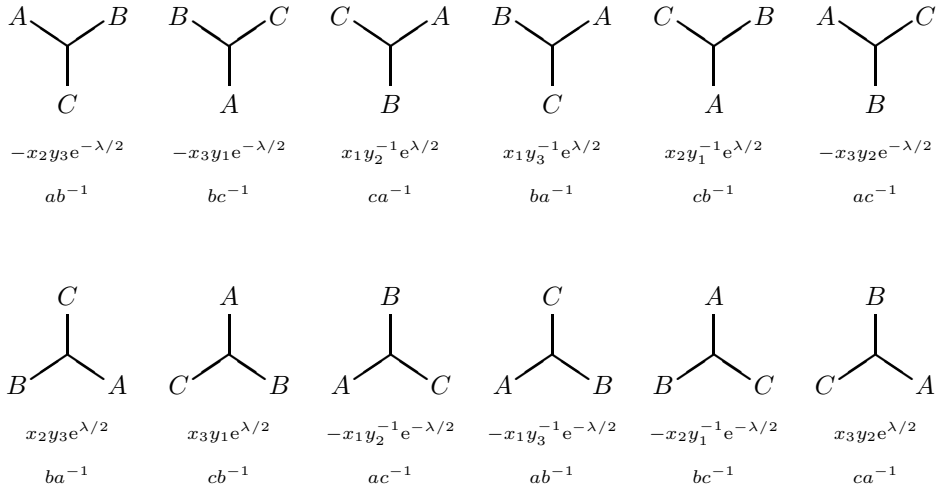

\centerline{\YHC{A}{B}{C}{-x_2y_3{\rm e}^{-\lambda/2}}{ab^{-1}}{.6mm}~\hspace{.3cm}~\YHC{B}{C}{A}{-x_3y_1{\rm e}^{-\lambda/2}}{bc^{-1}}{.6mm}~\hspace{.3cm}~\YHC{C}{A}{B}{x_1y_2^{-1}{\rm e}^{\lambda/2}}{ca^{-1}}{.6mm}~\hspace{.3cm}~\YHC{B}{A}{C}{x_1y_3^{-1}{\rm e}^{\lambda/2}}{ba^{-1}}{.6mm}~\hspace{.3cm}~\YHC{C}{B}{A}{x_2y_1^{-1}{\rm e}^{\lambda/2}}{cb^{-1}}{.6mm}~\hspace{.3cm}~\YHC{A}{C}{B}{-x_3y_2{\rm e}^{-\lambda/2}}{ac^{-1}}{.6mm}}\vspace{1cm}
\centerline{\laHC{B}{C}{A}{x_2y_3{\rm e}^{\lambda/2}}{ba^{-1}}{.6mm}~\hspace{.3cm}~\laHC{C}{A}{B}{x_3y_1{\rm e}^{\lambda/2}}{cb^{-1}}{.6mm}~\hspace{.3cm}~\laHC{A}{B}{C}{-x_1y_2^{-1}{\rm e}^{-\lambda/2}}{ac^{-1}}{.6mm}~\hspace{.3cm}~\laHC{A}{C}{B}{-x_1y_3^{-1}{\rm e}^{-\lambda/2}}{ab^{-1}}{.6mm}~\hspace{.3cm}~\laHC{B}{A}{C}{-x_2y_1^{-1}{\rm e}^{-\lambda/2}}{bc^{-1}}{.6mm}~\hspace{.3cm}~\laHC{C}{B}{A}{x_3y_2{\rm e}^{\lambda/2}}{ca^{-1}}{.6mm}}
\vskip5mm
\caption{Vertex configurations on the honeycomb lattice. On the
  first line their weights are given and on the second line the gauge
  transformations.}
\label{fig:HCweights}
\end{figure}

It follows that the model on the honeycomb lattice with vertex weights given in
Figure~\ref{fig:HCweights} (where $(123)$ $\rightarrow$ $(ABC)$) has
the same partition function as the model in Table~\ref{ta:BoltzW} at $u=-\lambda$ on the square lattice. More precisely,
\begin{equation}
\sinh(\lambda)^{-NM}Z_{NM}^{SQ}(u=-\lambda) = Z_{2NM}^{HC}(\lambda),
\label{eq:partsum}\end{equation}
The partition function does not change if we apply a gauge
transformation to the weights. If we
multiply the weights by the factors shown in Figure~\ref{fig:HCweights}
and choose $b=-{\rm e}^{-\lambda/3}a$ and $c={\rm e}^{-2\lambda/3}a$ and set
$x_i=y_i=1$, we obtain the partition sum of the fully packed loop (FPL)
model on the honeycomb lattice \cite{Reshetikhin:1989},
\begin{equation}
Z_{NM}^{SQ}(u=-\lambda) = \sinh (\lambda)^{NM}\sum_{\hat{C}}
n^{N(\hat{C})}.
\end{equation}
The sum runs over all dense loop coverings $\hat{C}$ of the honeycomb
lattice. $N(\hat{C})$ is the number of loops in the covering $\hat{C}$
and $n=2\cosh (\lambda)$ is the loop fugacity.

The edge states may be interpreted as differences modulo three (going
clockwise around each vertex) between three-state Potts variables on
the vertices of the triangular lattice. If the states $A$, $B$ and $C$
are interpreted as differences 0,1 and 2, the corresponding Potts
model allows only configurations on the triangle in which two
variables are equal and the third is different. The fully
ferromagnetic and the completely anti-ferromagnetic arrangements are
then excluded. This model has competing ferromagnetic two-spin and
anti-ferromagnetic three-spin interactions.

\section{Square-triangle tiling}
\begin{figure}[h]
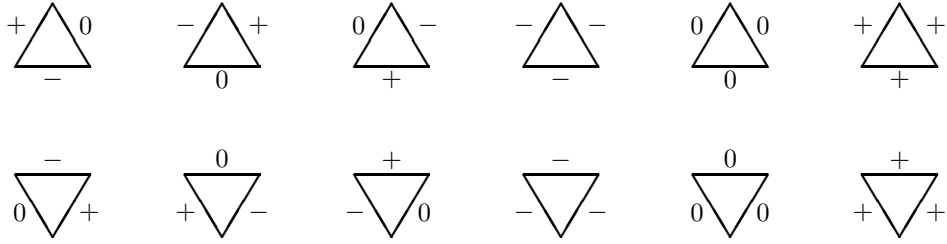

\centerline{\Tup{+}{0}{-}{1mm}~\hspace{8mm}\Tup{-}{+}{0}{1mm}~\hspace{8mm}\Tup{0}{-}{+}{1mm}~\hspace{8mm}\Tup{-}{-}{-}{1mm}~\hspace{8mm}\Tup{0}{0}{0}{1mm}~\hspace{8mm}\Tup{+}{+}{+}{1mm}}\vspace{8mm}
\centerline{\Tdown{0}{-}{+}{1mm}~\hspace{8mm}\Tdown{+}{0}{-}{1mm}~\hspace{8mm}\Tdown{-}{+}{0}{1mm}~\hspace{8mm}\Tdown{-}{-}{-}{1mm}~\hspace{8mm}\Tdown{0}{0}{0}{1mm}~\hspace{8mm}\Tdown{+}{+}{+}{1mm}}
\vskip5mm
\caption{Face configurations on the triangular lattice corresponding to the
  vertex configurations in Figure~\ref{fig:HCweights}}
\label{fig:Tweights}
\end{figure}

Take the dual of the honeycomb lattice and associate to each
edge of this triangular lattice the corresponding state variable of
the honeycomb lattice. Relabel the states of the horizontal axis by
$A$, $B$, $C$ $\rightarrow$ 0, +, $-$, on the ascending diagonal by
$A$, $B$, $C$ $\rightarrow$ +, $-$, 0 and on the descending diagonal by
$A$, $B$, $C$ $\rightarrow$ $-$, 0, +. The partition sum
(\ref{eq:partsum}) is then equal to the partition sum of a model on the
triangular lattice with face configurations given in
Figure~\ref{fig:Tweights}.
Now let $x_1^{-1}=x_2=x_3=x^{1/2}$ and $y_1=y_2=y_3=x^{-1/2}$.
If $x=0$ the faces with a 0 on all three edges vanish. In that case
the states $+$ and $-$ can be regarded as rotation angles of
edges w.r.t. a fixed triangular lattice. Take these angles $\pm
\pi /12$ and wipe out every edge with state 0.
In this way the model maps onto the square triangle random tiling
model. The state 0 corresponds to the diagonal of a square. 

The eigenvalue expression (\ref{eq:eigenval}) in the limit
$u=-\lambda$ becomes
\begin{eqnarray}
\Lambda(-\lambda) &=& x^{n_1+n_2}\sinh(-\lambda)^N
\prod_{k=1}^{n_1}\frac{\sinh(u_k^{(1)})}{\sinh
  (u_k^{(1)}+\lambda)} \prod_{l=1}^{n_2}
\frac{\sinh(u_l^{(2)}+2\lambda)}{\sinh (u_l^{(2)}+\lambda)}\nonumber\\
&&{} + x^{n_2}\sinh(-\lambda)^N \prod_{l=1}^{n_2}
\frac{\sinh(u_l^{(2)})}{\sinh(u_l^{(2)}+\lambda)}.
\end{eqnarray}
The Bethe Ansatz equations (\ref{eq:BAu2}) and (\ref{eq:BAu1}) for the
two sets of momenta become
\begin{equation}
\prod_{k=1}^{n_1}{\sinh (u_j^{(2)}-u_k^{(1)})\over \sinh
  (u_j^{(2)}-u_k^{(1)}+\lambda)}\prod_{l\neq j=1}^{n_2}{\sinh
  (u_j^{(2)}-u_l^{(2)}+\lambda) \over \sinh
  (u_j^{(2)}-u_l^{(2)}-\lambda)}=x^{n_1},\label{BAxu2}
\end{equation}
and
\begin{eqnarray}
\left({\sinh (u_j^{(1)}+\lambda)\over
  \sinh(u_j^{(1)})}\right)^N &=&
x^{N+n_2}\prod_{\stackrel{k=1}{\scriptscriptstyle k\neq
    j}}^{n_1}{\sinh (u_j^{(1)}-u_k^{(1)}+\lambda)\over \sinh
  (u_j^{(1)}-u_k^{(1)}-\lambda)}\times\nonumber\\ 
&&\prod_{l=1}^{n_2}{\sinh (u_j^{(1)}-u_l^{(2)}-\lambda)\over \sinh
  (u_j^{(1)}-u_l^{(2)})}.\label{BAxu1} 
\end{eqnarray}
It must be noted that these equations can also be derived from a
coordinate Bethe Ansatz. For the FPL model this was done by Baxter
\cite{Baxter:1970} whose method can be generalized slightly to obtain
the equations (\ref{BAxu2}) and (\ref{BAxu1}). For our purposes,
taking the limit $x\rightarrow 0$, it is more convenient to use
Baxter's variables. This can be accomplished by making the following
substitutions
\begin{equation}
s_j={x\sinh (u^{(1)}_j)\over \sinh (u^{(1)}_j+\lambda)},\qquad
t_\mu=-{x\sinh (u^{(2)}_\mu)\over \sinh (u^{(2)}_\mu+\lambda)}.
\end{equation}
On taking the limit $x\rightarrow 0$ we arrive at the following
expressions for the Bethe Ansatz equations and the eigenvalue of the
transfer matrix of the square-triangle tiling in its triangular
lattice representation 
\begin{eqnarray}
s_j^N = (-)^{n_1-1}\prod_{\nu=1}^{n_2} (s_j^{-1}-t_\nu^{-1}),\qquad
\prod_{k=1}^{n_1} (s_k^{-1}-t_\mu^{-1}) = (-)^{n_2-1}. \\
\Lambda = \left((-)^{n_2} + \prod_{k=1}^{n_1} s_k
\right)\prod_{\nu=1}^{n_2} t_\nu, 
\end{eqnarray}
These equations can
be solved analogously to the original solution of the
square-triangle random tiling by Kalugin \cite{Kalugin:1994}. 
\section{Conclusion}
We have made a connection between the recently solved square-triangle
random tiling model and a known solvable lattice model. It follows
from Table~\ref{ta:BoltzW} and the substitution $y_i=x^{-1/2}$ that
the $R$-matrix for the square-triangle tiling ($x=0$) is either
singular or contains infinite elements. As a result the transfer
matrix of the square-triangle tiling model is a limit of a family of
commuting transfer matrices, but is itself not a member of such family. For
any finite $x$ though the $R$-matrix is invertible, which implies
integrability via the Yang-Baxter equation for this more general model. The
square-triangle random tiling thus is a singular limit of a model
which is integrable in the usual sense. To obtain the weights of the
square-triangle model one has to take the limit $u \rightarrow
-\lambda$ first and then take $x \rightarrow 0$. These two limits do
not commute.
 
One final point can be made about the robustness of integrability. The
square-triangle tiling has been solved in three different ways. One is that of
Widom and Kalugin \cite{Widom:1993,Kalugin:1994}, the second can be
found in \cite{Gier:1996b} and the third one in this paper. All these
methods only differ in their choice of representation which of course
should not influence integrability. 
\section{Acknowledgment}
This research was supported by Stichting Fundamenteel Onderzoek der
Materie, which is part of the Dutch Foundation for Scientific
Research NWO.
\vfill\newpage
\appendix
\section{Diagonalization of the transfer matrix}
\label{diagT}
\begin{figure}[h]
\centerline{
\setlength{\unitlength}{1mm}
\begin{picture}(130,65)(-4,0)
\put(32,60){$F^{(0)}$}
\put(32,30){$F^{(0)}$}
\put(-4,52.5){$D:\; V_a^{(1)}$}
\put(-4,45){$B:\; V_2^{(1)}$}
\put(-4,37.5){$B:\; V_1^{(1)}$}
\put(60,52.5){$V_a^{(1)}$}
\put(60,45){$V_2^{(0)}$}
\put(60,37.5){$V_1^{(0)}$}
\put(72,45){=}
\put(64,11){$-$}
\put(10,0){\epsfig{file=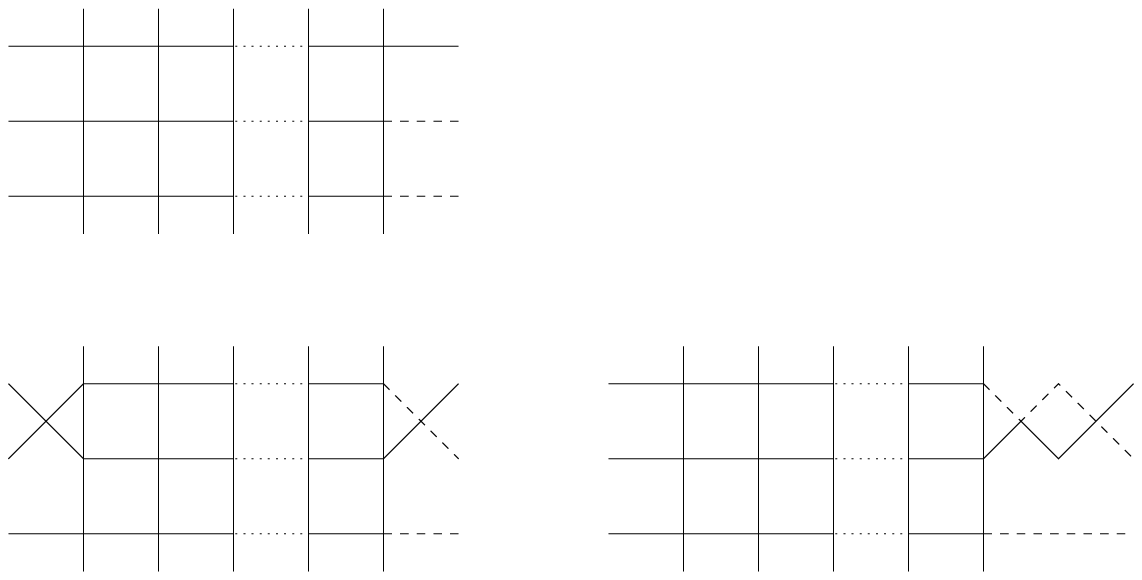}}
\end{picture}}
\vskip5mm
\caption{Commutation rule (\ref{commDB}) for $n_1=2$.}
\label{fig:commdiagram}
\end{figure}

We write the
transfer matrix $T(u)$ as an operator on
$(V^{(0)}\oplus V^{(1)})\otimes{\cal H}\simeq V\otimes{\cal H}$:
\begin{equation}
\left(\begin{array}{cc}
A(u) & B(u)\\
C(u) & D(u)
\end{array}\right)
\end{equation}
The entries of $T(u)$ act on the following spaces:
\begin{equation}
\begin{array}{ll}
A(u):\;V^{(0)}\otimes{\cal H}\rightarrow V^{(0)}\otimes{\cal
  H},& B(u):\;V^{(1)}\otimes{\cal
  H}\rightarrow V^{(0)}\otimes{\cal H}\\[2mm]
C(u):\;V^{(0)}\otimes{\cal H}\rightarrow V^{(1)}\otimes{\cal
  H},& D(u):\;V^{(1)}\otimes{\cal
  H}\rightarrow V^{(1)}\otimes{\cal H}.
\end{array}
\end{equation}
$V^{(0)}\simeq {\Bbb C}$ and $V^{(1)}\simeq {\Bbb C}^2$ are
the subspaces of $V$ corresponding to the natural embedding of
${\Bbb C}^2$ in ${\Bbb C}^3$. The $R$-matrix has the following form on the
standard basis of $(V_a^{(0)}\oplus V_a^{(1)})\otimes (V_b^{(0)}\oplus
V_b^{(1)})$,
\begin{equation}
R_{ab}(u) = \left(\setlength{\arraycolsep}{0pt}\begin{array}{@{}cccc}
\sinh (u+\lambda) & 0 & 0 & 0\\
0 & \sinh (u)U_b^{-1} & {\rm e}^{u}\sinh (\lambda)I^{(1)(0)} & 0\\
0 & {\rm e}^{-u}\sinh (\lambda)I^{(0)(1)} & \sinh (u)U_a & 0\\
0 & 0 & 0 & R_{ab}^{(1)}(u)
\end{array}\right).
\end{equation}
Here, $V_a^{(0)}\otimes V_b^{(1)} \stackrel{I^{(0)(1)}}{\rightarrow}
V_a^{(1)}\otimes V_b^{(0)} \stackrel{I^{(1)(0)}}{\rightarrow}
V_a^{(0)}\otimes V_b^{(1)}$ are canonical isomorphisms and
\begin{eqnarray}
U &=& \left(\begin{array}{cc}
y_3^2 & 0\\
0 & y_2^2
\end{array}\right).
\end{eqnarray}
With the notation $U_a$ we denote the operator that acts as $U$ in
the space $V^{(1)}_a$ and trivial everywhere else.
The reduced matrix $R_{ab}^{(1)}:\;V_a^{(1)}\otimes
V_b^{(1)}\rightarrow V_a^{(1)}\otimes V_b^{(1)}$ has the same
structure as the full $R$-matrix: 
\begin{equation}
R_{ab}^{(1)}(u) = \left(\setlength{\arraycolsep}{5pt}\begin{array}{@{}cccc}
\sinh (u+\lambda) & 0 & 0 & 0\\
0 & y_1^{-2}\sinh (u) & {\rm e}^{u}\sinh (\lambda) & 0\\
0 & {\rm e}^{-u}\sinh (\lambda) & y_1^2\sinh (u) & 0\\
0 & 0 & 0 & \sinh (u+\lambda)
\end{array}\right).
\end{equation}
From the relation (\ref{YBET}) all commutation relations between the
operators $A,B,C$ and $D$ can be obtained. In the sequel we shall only
need the following three of them,
\begin{eqnarray}
A_b(u) \otimes B_a(v) & = & \sinh (v-u)^{-1}\left(\vphantom{{\rm e}^{v-u}}\sinh
(v-u+\lambda) B_a(v) \otimes A_b(u)\right.\nonumber\\
&&\left.-{\rm e}^{v-u}\sinh (\lambda)B_b(u) \otimes
A_a(v)I^{(1)(0)}\right)U_a^{-1},\label{commAB}\\ 
D_a(v) \otimes B_b(u) & = & \sinh (v-u)^{-1}U_a^{-1}\left(B_b(u) \otimes D_a(v)
R^{(1)}_{ab}(v-u)\right.\nonumber\\
&&\left.-\vphantom{R^{(1)}_{ab}}{\rm e}^{u-v}\sinh (\lambda) I^{(0)(1)} B_a(v) \otimes
D_b(u)\right)\label{commDB}.\\
B_a(u) \otimes B_b(v) &=& \sinh(v-u+\lambda)^{-1} B_b(v) \otimes B_a(u) R_{ab}^{(1)}(v-u).   
\end{eqnarray}
The first step in diagonalizing ${\cal T}$ is to construct candidates
for its eigenvectors. This construction will be outlined below. One
first defines a 'pseudo-vacuum' $F^{(0)}$ on which $T(u)$ is
upper trigonal,
\begin{equation}
F^{(0)}=\bigotimes_{j=1}^N e_1,\;\;e_1= \left(\begin{array}{c} 1\\ 0 \\
0\end{array}\right).
\end{equation}
It then follows that 
\begin{equation}
T_a(u)F^{(0)}=\left(\setlength{\arraycolsep}{-16pt}
\begin{array}{@{}ccc}
x_1^{2N}\sinh (u+\lambda)^NF^{(0)} & * & *\\
0 & (x_2^2y_3^2)^N\sinh (u)^N F^{(0)} & *\\
0 & 0 & (x_3^2y_2^2)^N\sinh (u)^N F^{(0)}\hphantom{x_3}
\end{array}\;\right)
\end{equation}
$F^{(0)}$ therefore is an eigenvector of ${\cal T}(u)$. To obtain more
eigenvectors we make the following Ansatz 
\begin{eqnarray}
F &=& B_{n_1}(u_{n_1}^{(1)})\otimes\dots \otimes B_1(u_1^{(1)})
F^{(1),n_1},\label{eq:Fdef}\\
F^{(1),n_1} &\in& V_{n_1}^{(1)}\cdots\otimes V_1^{(1)} \otimes
F^{(0)}.\nonumber 
\end{eqnarray}
The vectors $F^{(1),n_1}$ will be found later. Schematically, we can
represent the action of the transfer matrix on $F$ as an $(n_1+1) \times N$
lattice, see Figure~\ref{fig:commdiagram}. The state $F^{(0)}$ is
represented by $N$ vertical edges and the action of each of the
$B_i(u_i^{(1)})$ by a horizontal line. The action of $D(u)$ on the
resulting vector $F$ is given by the upper horizontal line. The
commutation rule (\ref{commDB}) can now be represented graphically by
shifting the upper line downwards using the YBE (\ref{YBET}).  
In a similar fashion we can get a graphical representation of
(\ref{commAB}).\\
The factors that arise after commutation are given in
Figure~\ref{fig:commdiagram} by the external vertices. For example,
the vertex on the left of the second diagram of Figure~\ref{fig:commdiagram}
corresponds to $R^{(1)}(u-u_1^{(1)})$ and the one on the right to
$\sinh(u-u_1^{(1)})^{-1}U_a^{-1}$.  
It follows from the relations (\ref{commAB}) and (\ref{commDB}) that 
\begin{eqnarray}
A(u)F &=& x_1^{2N}\sinh(u+\lambda)^N \prod_{k=1}^{n_1} {\sinh
  (u_k^{(1)}-u+\lambda) \over \sinh (u_k^{(1)}-u)} \times\nonumber\\ 
&&\hphantom{x_1^N\sinh} B_{n_1}(u_{n_1}^{(1)})\otimes\dots
 \otimes B_1(u_1^{(1)}) U_{n_1}^{-1}\otimes\cdots \otimes U_1^{-1}
 F^{(1),n_1}\nonumber\\[2mm] 
&& +\;\mbox{`unwanted terms'.}\label{eq:AonF}\\[4mm]
D_a(u)F &=& \prod_{k=1}^{n_1} \sinh (u-u_k^{(1)})^{-1}
B_{n_1}(u_{n_1}^{(1)})\otimes \dots\otimes B_1(u_1^{(1)})\times\nonumber\\
&& T_a^{(1)}(u;\{u_k^{(1)}\})F^{(1),n_1}+\;\mbox{`unwanted
  terms'.}\label{eq:DonF}
\end{eqnarray}
The reduced transfer matrix is given by
\begin{eqnarray}
T_a^{(1)}(u;\{u^{(1)}_k\}) &=& U_a^{-n_1} D^{\vphantom{n_1}}_a(u)
R^{(1)}_{a1}(u-u_1^{(1)})\dots R^{(1)}_{an_1}(u-u_{n_1}^{(1)}),\\ 
T_a^{(1)}(u;\{u^{(1)}_k\})&:&\circlearrowleft V_a^{(1)}\otimes
V_{n_1}^{(1)}\otimes\cdots V_{1}^{(1)}\otimes {\cal H}.\nonumber 
\end{eqnarray}
The `unwanted terms' are similar to (\ref{eq:AonF}) and
(\ref{eq:DonF}) but now one of the $B_j$ has $u$ instead of
$u_j^{(1)}$ as its argument. Provided that the
`unwanted terms' vanish, the vector $F$ will thus be an
eigenvector of ${\cal T}(u)=A(u)+{\rm Tr}_aD_a(u)$ with eigenvalue
$\Lambda(u)$, given by 
\begin{eqnarray}
\Lambda(u) &=& x_1^{2N}\sinh(u+\lambda)^N \prod_{k=1}^{n_1} {\sinh
  (u_k^{(1)}-u+\lambda) \over \sinh (u_k^{(1)}-u)}\mu(U)\nonumber\\
&&{}+\prod_{k=1}^{n_1}
\sinh (u-u_k^{(1)})^{-1}\Lambda^{(1)}(u).\label{lambda0}
\end{eqnarray}
$U_{n_1}^{-1}\otimes\cdots \otimes U_1^{-1}$ and ${\cal
  T}^{(1)}(u;\{u_k^{(1)}\})$ are diagonal with eigenvalues $\mu(U)$ and
$\Lambda^{(1)}(u)$ respectively. For the `unwanted terms', it is
easily seen that the terms in which $B_{n_1}(u_{n_1}^{(1)})$ is
replaced by $B_{n_1}(u)$ are (up to a common multiplicative factor) precisely
of the form (\ref{eq:AonF}) and (\ref{eq:DonF}) with $u$ and $u_{n_1}^{(1)}$
interchanged and the factor with $k=n_1$ omitted.
Using the commutation rule for $B(u)B(v)$ it can be
shown that the same also holds for the terms with $B_j(u_j^{(1)})$
replaced by $B_j(u)$. The `unwanted terms' will therefore cancel if
$\Lambda(u_j^{(1)})=0$, or 
\begin{eqnarray}
x_1^{2N}\mu(U)\sinh (u_j^{(1)}+\lambda)^N
\prod_{k=1}^{n_1}\sinh (u_j^{(1)}-u_k^{(1)}-\lambda)=
-\Lambda^{(1)}(u_j^{(1)}) \label{BA1} 
\end{eqnarray}
The problem of finding the eigenvectors $F^{(1),n_1}$ of $T^{(1)}$ is
completely analogous to the 
construction above. It follows from the commutation rule 
\begin{equation}
R^{(1)}_{ab}(v-u) D_a(v) \otimes D_b(u) = D_b(u) \otimes D_a(v)
R^{(1)}_{ab}(v-u) 
\end{equation}
and the fact that $R_{ab}^{(1)}$ has the same form as $R_{ab}$ that
$T^{(1)}$ obeys the YBE
\begin{eqnarray}
R_{ab}^{(1)}(v-u)\cdot\left[T^{(1)}_a(v;\{u_k^{(1)}\})\otimes
T^{(1)}_b(u;\{u_k^{(1)}\})\right]=\nonumber\\
\left[T^{(1)}_b(u;\{u_k^{(1)}\})\otimes
T^{(1)}_a(v;\{u_k^{(1)}\})\right]\cdot R_{ab}^{(1)}(v-u).
\end{eqnarray} 
Writing $T^{(1)}(u;\{u_k^{(1)}\})$ as
\begin{equation}
T^{(1)}(u;\{u_k^{(1)}\})=
\left(\begin{array}{cc}
a(u) & b(u)\\
c(u) & d(u)
\end{array}\right),
\end{equation}
since $R^{(1)}(u)$ has the same structure as $R(u)$, we deduce
from (\ref{commAB}) and (\ref{commDB}) that
\begin{eqnarray}
a(u) b(v) & = & \sinh (v-u)^{-1}\left(\vphantom{{\rm e}^{v-u}}\sinh (v-u+\lambda) b(v)
a(u)\right.\nonumber\\
&&\left.{}-{\rm e}^{v-u}\sinh (\lambda)b(u) a(v)\right)y_1^{-2},\label{commab}\\
d(v) b(u) & = & y_1^{-2}\sinh (v-u)^{-1}\left(\vphantom{{\rm e}^{u-v}}\sinh (v-u+\lambda)b(u) d(v)
\right.\nonumber\\
&&\left.{}-{\rm e}^{u-v}\sinh (\lambda) b(v) d(u)\right)\label{commdb},
\end{eqnarray}
where now $b(u)$ and $b(v)$ commute.
Defining the `pseudo-vacuum' $F^{(1)(0)}$ in $V_{n_1}^{(1)}\otimes\cdots\otimes
V_{1}^{(1)}$ as
\begin{equation}
F^{(1)(0)} = \bigotimes_{j=1}^{n_1}
e^{(1)}_1,\;\;e^{(1)}_1=\left(\begin{array}{c}
1 \\ 0 \end{array}\right),
\end{equation}
the eigenvectors of ${\cal T}^{(1)}(u)$ are given by
\begin{equation}
F^{(1),n_1}=b(u_1^{(2)})\dots b(u_{n_2}^{(2)}) F^{(1)(0)} \otimes
F^{(0)}. \label{eigvecT1}
\end{equation}
\begin{figure}[h]
\setlength{\unitlength}{1mm}
\centerline{
\begin{picture}(130,50)(0,-10)
\put(65,-5){$F^{(0)}$}
\put(27,8){$F^{(1)(0)}$}
\put(20,0){\epsfig{file=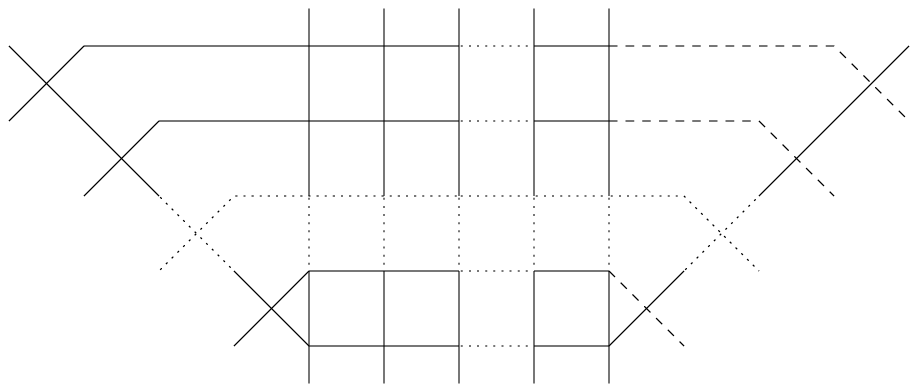}}
\end{picture}}
\caption{Action of ${\cal T}^{(1)}(u)$ on $F^{(1)(0)}\otimes F^{(0)}$.}
\label{fig:commdiagram2}
\end{figure}
Graphically, the action of ${\cal T}^{(1)}(u)$ on $F^{(1),n_1}$ is
depicted in Figure~\ref{fig:commdiagram2}. This diagram arises from
Figure~\ref{fig:commdiagram} when $D(u)$ is dragged down through all
$B$'s and all unwanted diagrams are discarded. The edges on the diagonal on the
left in Figure~\ref{fig:commdiagram2} now represent the `pseudo-vacuum'
$F^{(1)(0)}$ in the space $V_{n_1}^{(1)}\otimes\cdots\otimes
V_{1}^{(1)}$. The action of the $b_i$ on $F^{(1)(0)}$ is represented
by building a lattice on this diagonal in the same way as the
operators $B_i$ did on $F^{(0)}$, see
Figure~\ref{fig:commdiagram}. The commutation rules (\ref{commab}) and
(\ref{commdb}) can then be represented graphically similarly. 

The eigenvalues $\Lambda^{(1)}$ corresponding to the vectors
(\ref{eigvecT1}) are given by
\begin{eqnarray}
\Lambda^{(1)}(u)&=&y_3^{-2n_1}(x_2^2y_3^2)^Ny_1^{-2n_2}\sinh
(u)^N\prod_{k=1}^{n_1}\sinh (u-u_k^{(1)}+\lambda)\times\nonumber\\
&&\hphantom{{}+y_2^{-2n_1}(x_3^2y_2^2)^N}\prod_{l=1}^{n_2}{\sinh (u_l^{(2)}-u+\lambda)\over \sinh
  (u_l^{(2)}-u)}\nonumber\\  
&&{}+y_2^{-2n_1}(x_3^2y_2^2)^Ny_1^{2n_1-2n_2}\sinh
(u)^N\prod_{k=1}^{n_1}\sinh (u-u_k^{(1)})\times\nonumber\\
&&\hphantom{{}+y_2^{-2n_1}(x_3^2y_2^2)^N} \prod_{l=1}^{n_2}{\sinh (u-u_l^{(2)}+\lambda)\over \sinh
  (u-u_l^{(2)})}.\label{lambda1}
\end{eqnarray}
The eigenvalue $\mu(U)$ is simply given by
\begin{equation}
\mu(U)=y_3^{-2n_1+2n_2}y_2^{-2n_2}.
\end{equation} 
The `unwanted terms' generated by the action of $T^{(1)}(u)$ on
(\ref{eigvecT1}) can be read off from (\ref{commab}) and
(\ref{commdb}). Using the commutativity of $b(u)$ and $b(v)$ they can
be shown to cancel and make (\ref{eigvecT1}) an
eigenvector precisely when $\Lambda^{(1)}(u_k^{(2)})=0$. The numbers
$\{u_k^{(2)}\}$ therefore satisfy the equations
\begin{equation}
\prod_{k=1}^{n_1}{\sinh (u_j^{(2)}-u_k^{(1)})\over \sinh
  (u_j^{(2)}-u_k^{(1)}+\lambda)}\prod_{\stackrel{l=1}{\scriptscriptstyle l\neq j}}^{n_2}{\sinh
  (u_j^{(2)}-u_l^{(2)}+\lambda) \over \sinh
  (u_j^{(2)}-u_l^{(2)}-\lambda)}=\left({x_2^2y_3^2\over
  y_2^2x_3^2}\right)^N\left({y_2^2\over y_1^2y_3^2}\right)^{n_1}.
\label{eq:BAu2}\end{equation}
Knowing $\Lambda^{(1)}(u)$ the first set of equations as given by
(\ref{BA1}) becomes
\begin{eqnarray}
\left({\sinh (u_j^{(1)}+\lambda)\over
  \sinh(u_j^{(1)})}\right)^N &=& \left(x_2^2y_3^2\over
x_1^2\right)^N\left(y_2^2\over y_1^2y_3^2\right)^{n_2}\prod_{\stackrel{k=1}{\scriptscriptstyle k\neq j}}^{n_1}{\sinh
  (u_j^{(1)}-u_k^{(1)}+\lambda)\over \sinh
  (u_j^{(1)}-u_k^{(1)}-\lambda)}\times\nonumber\\
&&\hphantom{\left(x_2^2y_1^2\over x_1^2\right)^N}\prod_{l=1}^{n_2}{\sinh
  (u_j^{(1)}-u_l^{(2)}-\lambda)\over \sinh (u_j^{(1)}-u_l^{(2)})}  
\label{eq:BAu1}\end{eqnarray}
The eigenvalue combining the expressions (\ref{lambda0}) and
(\ref{lambda1}) becomes
\begin{eqnarray}
\Lambda(u) &=& x_1^{2N}y_3^{-2n_1+2n_2}y_2^{-2n_2}\sinh (u+\lambda)^N
\prod_{k=1}^{n_1} {\sinh (u-u_k^{(1)}-\lambda) \over \sinh
  (u-u_k^{(1)})}\nonumber\\
&&{}+\left(x_2^2y_3^2\right)^Ny_3^{-2n_1}y_1^{-2n_2}\sinh (u)^N
\prod_{k=1}^{n_1}{\sinh (u-u_k^{(1)}+\lambda) \over \sinh(u-u_k^{(1)})}
\times\nonumber\\
&&\hphantom{{}+\left(x_2^2y_3^2\right)^Ny_3^{-2n_1}}\prod_{l=1}^{n_2}{\sinh (u-u_l^{(2)}-\lambda)\over \sinh
  (u-u_l^{(2)})}\nonumber\\  
&&{}+\left(x_3^2y_2^2\right)^Ny_2^{-2n_1}y_1^{2n_1-2n_2}\sinh (u)^N
\prod_{l=1}^{n_2}{\sinh (u-u_l^{(2)}+\lambda)\over \sinh (u-u_l^{(2)})}.
\label{eq:eigenval}\end{eqnarray}

\end{document}